# Realization and characterization of single-frequency tunable 637.2 nm high-power laser

**Jieying Wang [1, 2], Jiandong Bai [1, 2], Jun He [1, 2, 3], and Junmin Wang [1, 2, 3]\***

[1]State Key Laboratory of Quantum Optics and Quantum Optics Devices (Shanxi University)

[2]Institute of Opto-Electronics, Shanxi University,

[3]Collaborative Innovation Center of Extreme Optics (Shanxi University),

No.92 Wu Cheng Road, Tai Yuan 030006, Shan Xi Province, People's Republic of China

**Abstract:** We report the generation of narrow-linewidth 637.2 nm laser by single-pass sum-frequency generation (SFG) of two infrared lasers at 1560.5 nm and 1076.9 nm in PPMgO:LN crystal. Over 8.75 W of single-frequency continuously tunable 637.2 nm laser is realized, and corresponding conversion efficiency is 38%. We study the behavior of crystals with different poling periods. The detailed experiments show that the output red lasers have very good power stability and beam quality. This high-performance 637.2 nm laser is significant for the realization of high power ultra-violet (UV) 318.6 nm laser via cavity-enhanced frequency doubling. Narrow-linewidth 318.6 nm laser is important for Rydberg excitation of cesium atoms via single-photon transition.

## 1. Introduction

Controlled interactions between parts of a many-body system is the key to quantum information processing. Neutral atoms represent promising approach [1-5]. The atoms in their ground state can store the quantum information, long-range interactions between highly excited Rydberg atoms are essential for successful operation of many quantum information protocols such as quantum network [6], entanglement [7] and implementation of elementary quantum gates [8]. For highly excited Rydberg state, due to the low transition probability and narrow linewidth, people usually choose a two-step or three-step excitation to a desired Rydberg state via the cascaded ladder-type transitions. However, multi-photon excitation will inevitably appear the population of intermediate state, which brings photon scattering and AC-Stark shift. Single-photon excitation can avoid the corresponding disadvantages. Generally, the wavelength of alkali metal atomic single-photon transition from ground state to the Rydberg state is in the UV band, so the excited laser needs to work in continuously tunable UV region, and with high power and narrow linewidth. As a challenging and promising excitation scheme, single-photon transition of Rydberg atoms has been demonstrated by Tong *et al*. [9] and Hankin *et al*. [10]. Based on the narrow-linewidth and continuously tunable fiber laser, we are intended to obtain the needed 318.6 nm UV laser from second harmonic generation (SHG) followed by sum-frequency generation (SFG) to drive the cesium-133 atoms from $6S_{1/2}$ state to $nP_{3/2}$ (n=80~100) Rydberg state via single-photon transition.

---

\* Corresponding author, email: *wwjjmm@sxu.edu.cn*

To generate the 318.6 nm UV laser, the frequency-doubling requires 637.2 nm red light as the fundamental laser. Unfortunately, the gain medium at 637.2 nm is very scarce. For example, the diode laser of this band is rare, and the output power is generally low, only a few mW to several tens mW, mostly used in laser pointer and laser collimation; A Ti:sapphire laser cannot be tuned to 637.2 nm; While the liquid dye laser can be tuned to 637.2 nm with specific dye, it can't work stably for a long time as a result of photo-bleach of the dye. Besides, the operation and maintenance of dye laser are very complex. So development of all solid state laser (such as from the nonlinear SFG), with good properties of operation and maintenance, has been an attractive approach. For example, some people like to get the high power quasi continuous-wave (CW) source of red-light from frequency doubling and sum frequency generation of all solid state laser [11,12].

Particularly, attribute to the quickly-developed laser technology such as fiber amplifier, narrow-linewidth fiber laser and quasi-phase-matched (QPM) nonlinear frequency conversion materials, many research groups continue to introduce new technology and better performance product [13-16]. 1.5 μm telecom Erbium-doped Fiber Amplifier (EDFA) and 1060-1080 nm band Ytterbium-doped Fiber Amplifier (YDFA) become more sophisticated, the output power can be up to dozens of Watts. In addition, with the fast development of the QPM nonlinear bulk crystals, the high-quality QPM crystals for the SHG, SFG, and parametric oscillator become available at acceptable price, such as periodically poled potassium titanyl phosphate (PPKTP) and periodically poled lithium niobate (PPLN), which have the advantage of large effective nonlinear coefficient ($d_{eff}$ (PPLN) ~17-18 pm/V and $d_{eff}$ (PPKTP) ~7-9 pm/V). For infrared band, people prefer to choose the PPLN crystal due to the higher $d_{eff}$. With the mature manufacturing technology, its interaction length can be up to 80 mm. Besides, the 5%-mol MgO-doped periodically poled lithium niobate (PPMgO:LN) can not only improve the photorefractive damage, but also reduce the phase-matching temperature, it's very suitable for using near the room temperature, which makes the application prospect of the PPLN crystals even better. Owning to the above technology, our group has made some investigations of singly resonant sum-frequency generation at 520 nm from 1560 nm and 780 nm laser, and 268 mW SFG output power was achieved [17]. For the great simplicity and stability, many people would like to prefer the single-pass configuration to obtain the continuous light output by the SFG. In 1999, Hart *et al.* demonstrated 0.45 W of 630 nm red laser output by single-pass SFG of an Nd:YAG laser and a 1.55 μm EDFA in a PPLN crystal [18]. In 2011, for the final aim of laser cooling $Be^+$ ions, 2 W tunable 626 nm red light was demonstrated via single-pass SFG by Wilson e*t al.* [13]. In 2014, Hankin *et al.* realized 1.1 W 638 nm laser output and firstly used it for the Rydberg excitation of cesium atoms via single-photon transition [10]. For the same purpose, we obtained far higher 637.2nm laser output, which provides a good experimental foundation for realization of high-power and narrow-linewidth 318.6 nm UV laser via frequency doubling.

In this paper，we describe a simple solid-state high power laser system which generated the 637.2 nm narrow-linewidth and continuously tunable laser. With the single-pass SFG configuration, up to 8.75 W single-frequency red laser is generated, by adding of two infrared laser at 1560.5 nm and 1076.9 nm within a 40-mm-long PPMgO:LN crystal, and the corresponding SFG conversion efficiency is up to 38.0%.

## 2. Experimental setup

A schematic of the experimental setup is shown in Fig. 1. Two commercial fiber lasers produced by NKT Photonics are used as seeding sources. The laser of 1560.5 nm is a distributed feedback Erbium-doped fiber laser (DFB-EDFL), with a nominal maximum output power of 200 mW, and a linewidth of <0.1 kHz. The wavelength tuning can be achieved by a relatively slow temperature coarse tuning with tuning range of 145.1 GHz and a fast piezo fine tuning with 3.7 GHz tuning range. The maximum output power of the high-power EDFA, with a wavelength operating range of 1540 nm to1565 nm, is about 15 W. The beam diameter of the EDFA is 1.4 mm. The other fiber laser is a distributed feedback Ytterbium-doped fiber laser (DFB-YDFL) at 1076.9 nm with a linewidth of 2 kHz, and a nominal maximum output power of 100 mW. The temperature tuning range is 202.3 GHz, and the piezo tuning range is 4.4 GHz. The maximum output power of the YDFA is 10 W, and its operating range is 1060 nm to1090 nm. The beam diameter is 1.7 mm.

Each beam after the amplifier passes through an optical isolator in order to prevent the optical feedback. A half-wave plate after each isolator is used to optimum the polarization state for the SFG. Two fundamental beams are overlapped by a dichroic mirror, and then focused into the QPM PPMgO:LN crystals. In this paper, two crystals with different poling periods are used to compare the SFG behavior. One of the PPMgO:LN crystal with a size of 30 mm×2 mm×1 mm and a poling period of 10.05 μm is manufactured by HC Photonics Corp. Another PPMgO:LN crystal with a size of 40 mm×10 mm×0.5 mm and five poling periods of 11.60, 11.65, 11.70, 11.75, and 11.80 μm is manufactured by Covesion Ltd.. In this experiment, we choose the period of 11.80 μm, which corresponds the lowest optimized QPM temperature. The input and output surface of both crystals are anti-reflection (R<0.2% for 30-mm-long crystal, R<1% for 40-mm-long crystal) coated for 1560.5, 1076.9, and 637.2 nm.

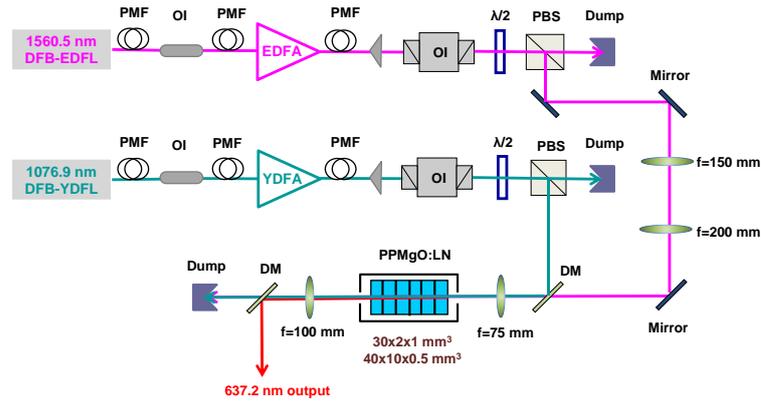

**Fig. 1.** Schematic diagram of the single-pass SFG system. Infrared light from two fiber lasers are frequency summed in PPMgO:LN to produce 637.2 nm red light. DFB-EDFL: distributed feedback Erbium-doped fiber laser, EDFA: Erbium-doped fiber amplifier, DFB-YDFL: distributed feedback Ytterbium-doped fiber laser, YDFA: Ytterbium-doped fiber amplifier, PMF: polarization-maintaining optical fiber, OI: optical isolator, λ/2: half-wave plate, PBS: polarization beam splitter cube, DM: dichromatic mirror, Mirror: 45º high-reflectivity mirror.

## 3. Experimental results and discussion

Fig. 2 shows the temperature tuning curves of both the QPM-crystals. (a) is the result of 30-mm-long crystal, both of the input fundamental powers are fixed at 2 W. The highest 637.2 nm SFG power is obtained when the crystal temperature is tuning to 63.0 ℃. (b) is for the 40-mm-long crystal. The measured QPM temperature is 154 ℃ when both of the fundamental powers are 1 W. The full-width of half-maximum (FWHM) bandwidth of the phase-matching temperatures are 1.5 ℃ and 1.2 ℃, respectively. The phase-matching temperature bandwidth can be given as [19]:

$$\Delta T \cdot l = \frac{2\lambda_1}{2.25} \left| \frac{\partial n_1}{\partial T} + \frac{\lambda_1}{\lambda_2} \frac{\partial n_2}{\partial T} - \frac{\lambda_1}{\lambda_3} \frac{\partial n_3}{\partial T} \right|^{-1} \tag{1}$$

Here, ΔT indicates the temperature bandwidth, $l$ is the crystal length, $\lambda_i$ (i =1, 2, 3) denote the fundamental (index1, 2) and sum-frequency (index3) wavelength, $n_i$ is the refractive index in the crystal, and $\partial n_i/\partial T$ represent the temperature coefficients of the fundamental and sum-frequency components within the crystal. For a certain nonlinear interaction, when phase matching conditions is reached, Eq. (1) give that the temperature bandwidth is inversely proportional to the crystal length. Our experimental results verify it, the temperature tuning bandwidth of 30-mm-long crystal is wider than the 40-mm-long crystal.

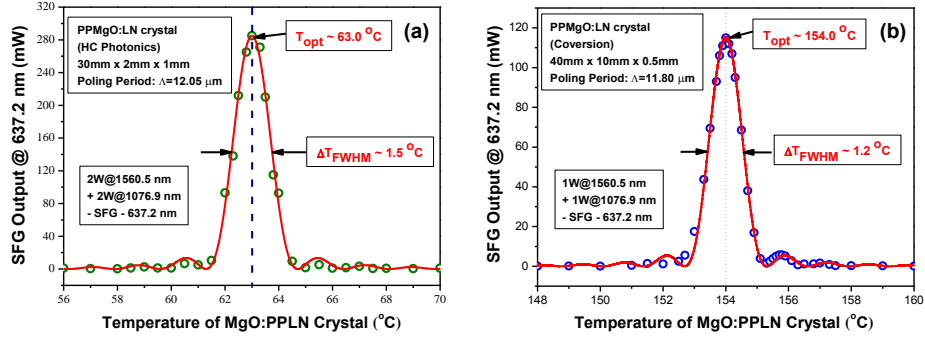

**Fig. 2.** The temperature tuning curves of the PPMgO:LN crystals for single-pass sum-frequency generation. Circles are the experimental data, while the solid lines are the theoretically fitted curves using $sinc^2$ function. (a) PPMgO:LN crystal of the dimensions 30 mm×2 mm×1 mm (Poling Period: Λ=12.05 μm), and the QPM optimized temperature is 63 ℃ with a FWHM of 1.5 ℃; (b) PPMgO:LN crystal of the dimensions 40 mm×10 mm×0.5 mm (Poling Period: Λ=11.80 μm), and the QPM optimized temperature is 154 ℃ with a FWHM of 1.2 ℃.

In order to achieve the maximum SFG efficiency, the spatial mode matching of the two fundamental beams within the crystal is the most important. However, choosing appropriate size of beam waist to make full use of the whole crystal's length also should be taken into account. So the optimum SFG can be obtained when the confocal parameters of these two fundamental beams are equal and the focusing parameters $\xi=l/b=2.84$, where $b=2\pi\omega_0^2/\lambda$ [20]. Integrating these factors and in view of simplifying the experimental system, the two fundamental beams are focused into the PPMgO:LN by a 75mm lens. The lens produced two beam waists of 43 μm ($w_{1560\ nm}$) and 30 μm ($w_{1077\ nm}$).

The 637.2 nm SFG power versus the incident power of 1560.5 nm laser is shown in Fig. 3. In the process of measurement, the power of 1076.9 nm is fixed at 9 W and the power of 1560.5 nm varied. To achieve the maximum sum-frequency conversion, the QPM temperature must be optimized at each combination of input power. For the two different crystals, the maximum 637.2 nm red light of 7.87 W (30-mm-long crystal) and 8.75 W (40-mm-long crystal) are generated when the two fundamental lasers power are 9 W and 14 W, these conditions correspond to an optical to optical conversion efficiency ($\frac{P_3}{P_1+P_2}$) of 34.2% and 38.0%, respectively. Fitting to the data of the linear region yield the nonlinear conversion efficiency of 2.74% (W cm)$^{-1}$ and 2.36%

(W cm)$^{-1}$, respectively. The SFG efficiency can be written as [18]:

$$\eta = \frac{P_3}{P_1 P_2 l} = \frac{8\omega_0^3 d_{eff}^2}{\pi\varepsilon_0 c^4 n_0 n_3} \left[\frac{(1-\delta^2)(1-\gamma^2)}{(1+\delta\gamma)}\right] h(\mu,\xi) \sin^2(\pi\Delta) \quad (2)$$

Here, $P_i$ (i =1, 2, 3) represent the laser power of 1560.5, 1076.9 and 637.2 nm, respectively. $n_o=(n_1+n_2)/2$, $\omega_0=(\omega_1+\omega_2)/2$, $\omega_i$ is the frequency of the corresponding laser, $d_{eff}$ is the effective nonlinear coefficient. $\delta=1-2\omega_1/(\omega_1+\omega_2)$ and $\gamma=1-2n_1/(n_1+n_2)$, $h(\mu, \xi)$ indicates the B-K focusing factor, it's related to the focusing parameter $\xi$. The last term in Eq. (2) is the correct for the crystal's non-ideal grating duty cycle.

For the case of 30-mm-long crystal, the focusing parameter $h(\mu, \xi)$ is estimated to be 1.0, and the nonlinear conversion efficiency is 3.7% (W cm)$^{-1}$. For the 40-mm-long crystal, $h(\mu, \xi)$ is estimated to be 0.9, the corresponding conversion efficiency is 3.4% (W cm)$^{-1}$. Because our focusing condition is more appropriate for the 30-mm-long crystal, so theoretical calculation and experimental results show that the nonlinear conversion efficiency of 30-mm-long crystal is larger than the 40-mm-long crystal.

Possible reasons for the discrepancy between theory and experiment are as follows: 1) Beam clipping and diffraction loss at the edges of the relatively thin PPMgO:LN crystal, this phenomenon is especially obvious in the 40-mm-long crystal; 2) The single mode-matching lens introduce the chromatic aberration, and the distance between the two waist of the fundamental beams is about 500 μm, so the space mode of this two fundamental lasers can not overlap perfectly; 3) The uneven temperature profile inside the crystal makes the whole temperature within the crystal can't simultaneously reach the optimum phase-matching point; 4) The fabrication defects of crystals.

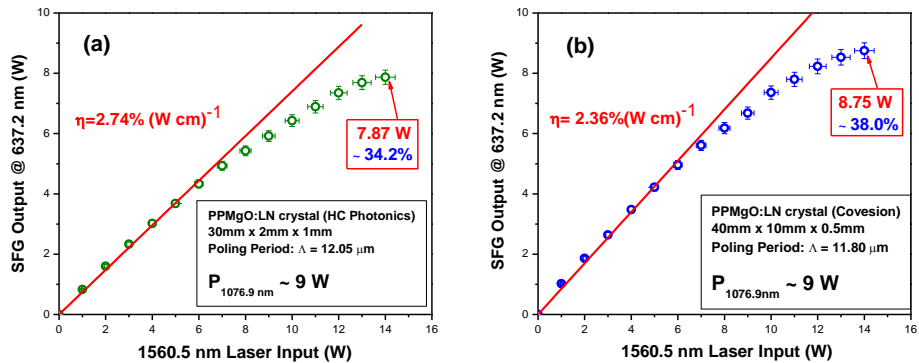

**Fig. 3.** SFG output power versus power of two fundamental lasers using a 75 mm focusing lens. The 1076.9 nm laser power was fixed at 9 W and the 1560.5 nm laser power varied. The error bars come from the measurement error of power meter. (a) the case of 30-mm-long PPMgO:LN crystal from HC Photonics; (b) the case of 40-mm-long PPMgO:LN crystal from Covesion.

Since the SFG process satisfy the conservation of energy, one 1560.5 nm photon and one 1076.9 nm photon are converted to one 637.2 nm photon. According to the consumption of photons, the efficient SFG conversion efficiency can be obtained when the ratio of power between 1560.5 nm and 1076.9 nm is equal to their ratio of frequency. Calculated power ratio for 1076.9 nm and 1560.5 nm is 9:6.2. However, it's hard for the two fundamental Gaussian beams to reach the completely spatial mode-matching, and there is maybe absorption-induced heating in the PPMgO:LN crystals in the high power region [21]. So the SFG process can't completely follow the consumption ratio of photons, and from Fig. 3 we just observed the phenomenon of deviation from linear after the power of 1560.5 nm above 6 W, not fully saturated. If a periodically poled waveguide is adopted, the space mode of two fundamental beams can approximate to achieve a good match, so the two fundamental lasers' power consumption ratio can be much closer to the theoretical value [22], but this approach is not suitable for high power conditions.

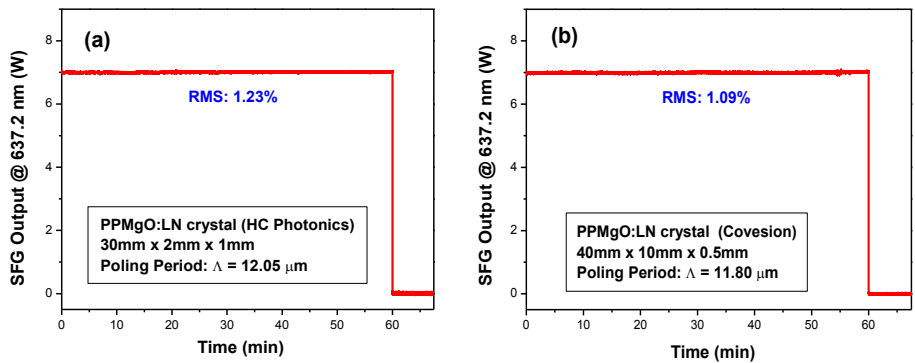

**Fig. 4.** Power stability of the SFG output over 1 hour. The RMS fluctuations are 1.23% in the case of 30-mm-long PPMgO:LN crystal (a) and 1.09% in the case of 40-mm-long PPMgO:LN crystal (b) under the output power around 7W respectively.

The output 637.2 nm SFG power is very Stable. We monitored the stability of 7 W SFG power over 1 hour, and the typical results are shown in Fig.4, (a) is for the 30-mm-long PPMgO:LN crystal, the root-mean-square (RMS) fluctuation is less than 1.23%; (b) is for the 40-mm-long PPMgO:LN crystal, it is less than 1.09%. This fluctuation observed is mainly due to the slow changes in polarization from the fiber lasers and the fiber amplifiers. These changes should be attributed to the small temperature variation and the air disturbance of laboratory. Besides, the crystal's absorption to the high-power fundamental lasers heats itself in the core region, which leads to the temperature drift in the crystals. Therefore, the temperature controllers' resolution and the thermal contact between the crystal and the oven became very important. In the case of 30-mm-long crystal, the crystal oven is home-made which is heated by a temperature controller (Newport, model 350B). While in the other case, it is a Covesion oven (model PV40) and its own heater controller (model OC1). The resolution of both temperature controllers are better than 0.05 ℃, which is adequate for the FWHM of 1.5 ℃ and 1.2 ℃, respectively. However, the thermal contact surface of 40 mm×10 mm×0.5 mm crystal is much larger than the 30 mm×2 mm×1 mm crystal, so the power stability of 40-mm-long crystal is better.

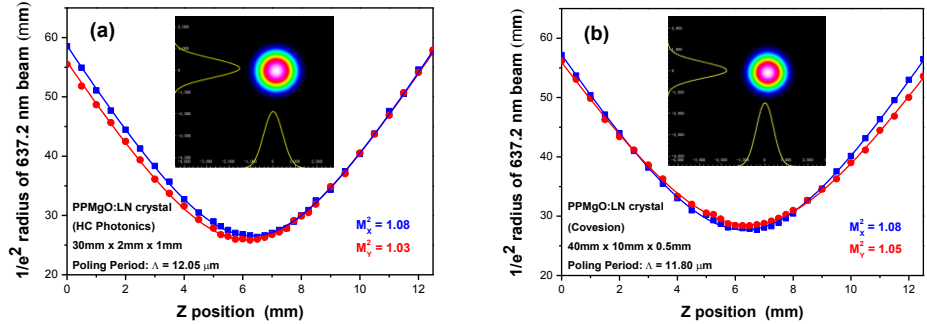

**Fig. 5.** $M^2$ factors of the SFG beam. The blue squares show the measurements of the horizontal direction X, and the red circles show the vertical direction Y. Insets show the typical intensity profile of the SFG laser beam. (a) the case of 30-mm-long PPMgO:LN crystal; (b) the case of 40-mm-long PPMgO:LN crystal.

The beam propagation parameters of the generated 637.2 nm laser are fitted by measuring the beam quality $M^2$ in two orthogonal transverse directions X and Y, as shown in Fig. 5. (a) shows the result of the 30-mm-long crystal, the measured $M^2$ of both directions are 1.08 and 1.03, respectively. (b) shows the result of the 40-mm-long crystal, they are 1.08 and 1.05, respectively. The $M^2$ are close to 1, and confirming $TEM_{00}$ spatial mode. This is mainly attributed to the optical fiber output beam of the fundamental fiber amplifiers, whose beam qualities are very well.

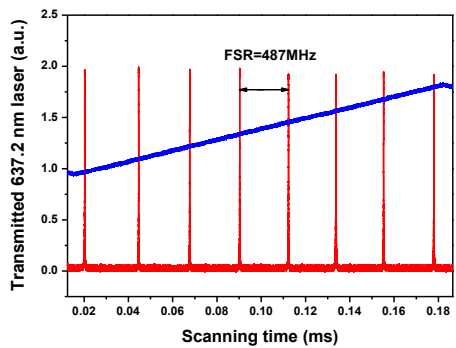

**Fig. 6.** Scanning 637.2 nm laser output monitored by an optical cavity with a free spectral range (FSR) of 487 MHz.

**8 / 11**

We analyze the transmitted spectrum of the generated 637.2 nm laser using a monitor cavity, with a free spectral range (FSR) of 487 MHz. While the 1076.9nm fiber laser is slowly swept, a typical fringe pattern shown in Fig. 6 is obtained. The red laser can be smoothly tuned across more than 7FSRs of the monitor cavity, indicates the continuously tunable range of the 637.2 nm laser's frequency is ~ 3.4 GHz, which is limited by the fiber laser at 1076.9 nm.

Since the lifetime of the cesium atoms $nP_{3/2}$ (n=80~100) is about 270 μs, the estimated $6S_{1/2}$-$nP_{3/2}$ (n=80~100) transition linewidth is about 5.9 kHz. In order to carry out the single-photon Rydberg excitation, the 318.6 nm laser's linewidth must be narrow. Our narrow linewidth DFB fiber lasers create the possibility. We estimate the linewidth of the 637.2 nm laser using an ultralow expansion high-finesse cavity (FSR=3.145 GHz, Finesse > 30000) and can place a upper bound of 105 kHz. Besides, considering that the nominal linewidth of 1560.5 nm DFB-EDFL and 1076.9nm DFB-YDFL are 0.1 kHz and 2 kHz, and the EDFA and YDFA have selected the option to keep the narrow linewidth, so the linewidth can't be obviously broadened in the process of SFG, we estimate the linewidth of the 637.2 nm red light should be less than 10 kHz.

## 4. Conclusion

In conclusion, we have demonstrated a continuous-wave tunable single-frequency 637.2 nm laser using single-pass sum-frequency approach. 8.75 W of 637.2 nm laser has been obtained by combining of 9 W from a 1076.9 nm YDFA and 14 W from a 1560.5 nm EDFA within the 40-mm-long PPMgO:LN crystal, and the corresponding maximum sum-frequency conversion efficiency is 38.0%. We have studied two different poling periods of PPMgO:LN bulk crystals where the generated 637.2 nm laser performed a very well power stability, and the beam quality is nearly perfect $TEM_{00}$ mode. The Watt-level red laser has its wide application in the field of large area projection systems, laser therapy, and laser ranging. Especially, 637.2 nm laser provides a good experimental foundation for realization of high-power and narrow-linewidth 318.6 nm UV laser via frequency doubling. The 318.6 nm UV laser can be used to achieve the single-photon Rydberg excitation of cesium atoms.

**Acknowledgments:** This project is supported by the National Natural Science Foundation of China (61475091, 11274213, 61205215, and 612297902) and the National Major Scientific Research Program of China (2012CB921601).

of sodium resonance radiation at 589 nm by using a periodically poled Zn:LiNbO$_3$ ride waveguide, Opt. Express 17 (2009) 17792-17800.